\documentclass[useAMS,usenatbib]{mn2e}
\usepackage{epsfig}
\bibpunct{(}{)}{;}{a}{,}{,}
\usepackage[usenames,dvipsnames]{color}
\usepackage{lscape}
\usepackage{rotating}
\usepackage{placeins}

\def\gtrsim{\mathrel{\hbox{\rlap{\hbox{\lower4pt\hbox{$\sim$}}}\hbox{$>$}}}}
\def\lesssim{\mathrel{\hbox{\rlap{\hbox{\lower4pt\hbox{$\sim$}}}\hbox{$<$}}}}

\title[The composition of UCDs]{The chemical composition of Ultracompact Dwarf
  Galaxies in the Virgo and Fornax Clusters}

\author[Francis et al.]{K.J.\
  Francis$^{1}$\thanks{E-mail: katy.francis@uqconnect.edu.au}, M.J.\
Drinkwater$^{1}$, Igor V.\ 
Chilingarian$^{2,3}$, A.M. \ Bolt$^{1}$, P.\ Firth$^{1}$
\\
$^{1}$School of Mathematics and Physics, The University of Queensland,
Brisbane, QLD 4072, Australia\\
$^{2}$Smithsonian Astrophysical Observatory, Harvard-Smithsonian Center for 
Astrophysics, Cambridge, MA 02138, USA \\
$^{3}$Sternberg Astronomical Institute, Moscow State University, 13
Universitetski prospect, 119992, Moscow, Russia }

\begin{document}


\pagerange{\pageref{firstpage}--\pageref{lastpage}} \pubyear{2009}

\maketitle

\label{firstpage}

\begin{abstract}
 We present spectroscopic observations of ultra compact dwarf (UCD) galaxies in the Fornax and Virgo Clusters made to measure and compare their stellar populations. The spectra were obtained on the Gemini-North (Virgo) and Gemini-South (Fornax) Telescopes using the respective Gemini Multi-Object Spectrographs. 

We estimated the ages, metallicities and abundances of the objects from measurements of Lick line-strength indices in the spectra; we also estimated the ages and metallicities independently using a direct spectral fitting technique. Both methods revealed that the UCDs are old (mean age $10.8\pm 0.7$ Gyr) and (generally) metal-rich (mean [Fe/H] = $-0.8\pm 0.1$). The alpha-element abundances of the objects measured from the Lick indices are super-Solar. 

We used these measurements to test the hypothesis that UCDs are formed by the tidal disruption of present-day nucleated dwarf elliptical galaxies. The data are not consistent with this hypothesis because both the ages and abundances are significantly higher than those of observed dwarf galaxy nuclei (this does not exclude disruption of an earlier generation of dwarf galaxies). They are more consistent with the properties of globular star clusters, although at higher mean metallicity. The UCDs display a very wide range of metallicity ($-1.7<$[Fe/H]$<0.0$), spanning the full range of both globular clusters and dwarf galaxy nuclei.

We confirm previous reports that most UCDs have high metalliticities for their luminosities, lying significantly above the canonical metallicitiy-luminosity relation followed by early-type galaxies.
In contrast to previous work we find that there is no significant difference in either the mean ages or the mean metallicities of the Virgo and Fornax UCD populations. 
\end{abstract}


\section{Introduction}


Ultracompact dwarf (UCD) galaxies were originally discovered as very
compact stellar systems (CSS) in the nearby Fornax galaxy
cluster \citep{Hilker1999,Drinkwater2000}. Although as luminous as
dwarf galaxies ($-11<M_B<-12$), these objects were easily confused
with Galactic stars in ground-based imaging. Measurements of the
structural properties of these objects established that they were
distinct from both globular clusters (more luminous and more distant
from galaxies), and dwarf galaxies (much more
compact) \citep{Phillipps2001,Drinkwater2003}.  UCDs have since been
found in the Virgo cluster \citep{Jones2006_VUCD} and other more
distant clusters \citep[see][for examples]{Baumgardt2008}.

The location of the UCD populations in dense galaxy clusters close to
central galaxies with rich globular cluster populations prompted two
main hypotheses for their formation. One model is that UCDs formed when the
outer layers of nucleated dwarf galaxies are removed by tidal
stripping near the cluster centre
\citep{Bekki2001_tidal, Bekki2003_tidal,
  Thomas2008, Paudel2010}.  The other model is that UCDs are simply
the extremely bright tail of the usual globular cluster (GC)
populations around the central galaxies in each cluster 
  \citep{Evstigneeva2007_VUCD, Mieske2012}. It is not as easy as
first   thought to separate these two models, so the question of UCD
formation remains a topic of active investigation 
  \citep[e.g.][]{Chilingarian2011, Brodie2011_new, Chiboucas2011,
    Bruns2011}. 

Many studies have measured the structural properties of UCDs,
benefiting in particular from {\em Hubble Space Telescope} imaging to
resolve the objects. The latest analyses of the internal dynamics (as
well as their stellar mass functions) show no need for dark matter in
these objects to explain their mass-to-light
ratios \citep{Baumgardt2008, Chilingarian2011} in contrast to some
earlier results. On the other hand, there is increasing evidence that
the more massive UCDs exhibit structural relationships unlike
GCs. UCDs more luminous than $M_V \approx -11$ display a strong
correlation between size and luminosity, whereas the sizes of globular
clusters are not correlated with
luminosity \citet{Mieske2006_UCD,Evstigneeva2008}.

The other major observational property of UCDs---the focus of this
paper---is their chemical composition. The availability of 8-m class
telescopes makes it quite feasible to make spectroscopic estimates of
the ages and metallicities of UCDs in the nearer clusters. \citet{Mieske2006_UCD} obtained
Magellan Telescope spectra of 26 compact Fornax objects: the bright
UCDs had high metallicities compared to dE galaxies in Fornax,
suggesting they were very young, although the ages were not measured
directly. In contrast Keck spectra of a sample of Virgo cluster
UCDs \citep{Evstigneeva2007_VUCD} showed they were all old (more than 8 Gyr)
like globular clusters. More evidence for a difference between the two
clusters was presented by \citet{Firth2009_UCD} who also reported
young ages for some Fornax cluster UCDs, although these were measured
with lower quality spectra. More recently, \citet{Chilingarian2011}
analysed VLT spectra of 24 Fornax UCDs finding old ages consistent
with globular clusters.


In this paper we use high-quality data to make accurate measurements
of the chemical composition of UCDs in both the Fornax and Virgo
clusters. Our aim is to provide further tests of the formation models
discussed above. In particular we wish to test the previous claims
that the Fornax UCDs are younger than those in Virgo. We will do this
by using data from similar instruments and identical analysis for both
clusters to minimise any systematic effects.

We describe our observational data and spectral analysis in
Section \ref{sec-obs}. This consists of high signal-to-noise long slit
spectra of Fornax and Virgo UCDs obtained with the Gemini North and
South Multi-Object Spectrographs (GMOS-N, and GMOS-S). In
Section \ref{sec-analysis} we compare the results to stellar
population models to determine the ages, metallicities, and
$\alpha$-element abundances of the UCDs. We discuss the implications of
the measurements in Section \ref{sub-theories} and we summarise the paper
in Section \ref{sec-summary}.

\section{Observations and Data Reduction}
\label{sec-obs}

\subsection{Selection of Targets}
\label{sub-targets}

We selected targets for our Gemini observations from our larger UCD
samples in the Fornax \citep{Gregg2009_UCD} and
Virgo \citep{Jones2006_VUCD,Firth2008_paper} clusters. The targets
were selected at random from the central (1 degree radius
corresponding to a projected separation from the central galaxy of
$200$ kpc) field in each cluster, chosen so as to have luminosities in
the range $-10<M_B<-12$. The only exception to this selection was that
two Virgo objects from the ``Intra-cluster field'' were included in
the Virgo sample. The mean luminosities of the targets were $M_B =
-11.0 \pm 0.2$ and $-11.1 \pm 0.2$ for the Fornax and Virgo objects
respectively\footnote{We quote the standard error
    of the mean for each value}. There were no significant 
differences in the luminosity 
distributions in the two clusters.  The UCDs measured are listed in
Table~\ref{tab-vel} along with one globular cluster in each region
observed for comparison purposes. The table lists names for the
objects based on their coordinates except when names were previously
assigned by \citep{Drinkwater2000} or \citep{Jones2006_VUCD}.


For absolute magnitudes, we converted the SDSS $g$ magnitudes to Cousins $B$ magnitudes using the relation $B =g +0.42(g-r)  - 0.024$ \citep{Fukugita1996}.

We adopt the following distance moduli in this paper: the Fornax
Cluster at 31.35 mag, the Virgo Cluster at 30.97 mag \citep[see Table
1 of][]{Firth2008_paper} and the comparison globular cluster in NGC
1407 at 31.99 mag \citep{Jerjen2004}.  


\subsection{Gemini Observations}

The spectroscopic data for ultra-compact dwarf galaxies were collected using
GMOS-N and GMOS-S spectrographs on the 8-m GEMINI-North and GEMINI-South
telescopes for Virgo and Fornax objects correspondingly. All observations
were obtained in the service mode by the GEMINI observatory staff. Both
spectrographs are equipped with 6.3k$\times$4.5k mosaic detectors comprising
three 2k$\times$4.5k CCD chips with narrow 90-pixel gaps between
them.  We describe our processing of the two sets of data separately
due to small differences in the setup. Data from one of the observing
runs (GN-2007A-Q76) were previously published \citep{Firth2009_UCD},
but we reprocessed those data for the current paper to obtain a
uniform set of measurements.

\begin{table*}
\caption{Summary of the Targets }
\label{tab-vel}
\begin{tabular}{lclc r@{:}c@{:}l r@{:}c@{:}l r@{ $\pm$ }l c c c}
\hline
Name(s) & Run$^{a}$  & Slit Width & T$_{exp}^{b}$ &
\multicolumn{3}{c}{RA} & \multicolumn{3}{c}{Dec} & \multicolumn{2}{c}{Velocity} &
\multicolumn{2}{c}{Magnitude} & Radius$^{c}$\\ 
 &  & (arcsec) & (s) & \multicolumn{3}{c}{(h:m:s)} &
 \multicolumn{3}{c}{ ($^{\circ}$:$'$:$"$) } &\multicolumn{2}{c}{ (km
   s$^{-1}$) } &
 \multicolumn{2}{c}{(mag)} & (arcmin) \\ 
\hline
 & & & &\multicolumn{3}{c}{ }  & \multicolumn{3}{c}{ } &
 \multicolumn{2}{c}{ } & $g^{d}$ & $r$ $^{d}$ & \\ 
FUCD0336-3536 & 3 & 0.75 & 4800 & 03&36&22.28 & -35&36&34.3 & 1297&22
& 20.29 & 19.77 & 27.50 \\ 
FUCD0336-3522 & 3 & 0.75 & 4800 & 03&36&26.72 & -35&22&01.6 & 1290&45
& 20.20 & 19.42 & 25.41  \\ 
FUCD0336-3514 & 3 & 0.75 & 4200 & 03&36&27.74 & -35&14&13.9 & 1416&14
& 20.12 & 19.27 & 27.84  \\ 
FUCD0336-3548 & 3 & 0.75 & 7500 & 03&36&47.74 & -35&48&34.1 & 1416&18
& 20.74 & 20.04 & 29.80  \\ 
UCD1          & 3 & 0.75 & 3000 & 03&37&03.30 & -35&38&04.6 &  1563&14
& 19.79 & 19.03  & 20.65  \\ 
FUCD0337-3536 & 3 & 0.75 & 4800 & 03&37&24.91 & -35&36&09.7 & 1502&35
& 20.26 & 19.61 & 15.93  \\ 
FUCD0337-3515 & 3 & 0.75 & 7500 & 03&37&43.56 & -35&15&09.6 & 1283&21
& 20.79 & 20.12 & 15.05  \\ 
FUCD0338-3513 & 3 & 0.75 & 4200 & 03&38&23.75 & -35&13&49.5 &  1565&25
& 20.13 & 19.57 & 13.23  \\ 
FUCD0339-3519 & 3 & 0.75 & 4200 & 03&39&20.56 & -35&19&14.6 &  1466&13
& 20.17 & 19.42 & 13.07  \\ 
UCD5          & 3 & 0.75 & 3000 & 03&39&52.58 & -35&04&24.1 & 1282&18
& 19.86 & 19.25 & 28.33  \\ 
NGC1407-GC1   & 3 & 0.75 & 7500 & 03&40&09.42 & -18&33&37.3 & 1219&27
& $B=20.98^{e}$ & $I=19.16^{e}$ & 1.33  \\ 
\hline
 & & & &\multicolumn{3}{c}{ }  & \multicolumn{3}{c}{ } &
 \multicolumn{2}{c}{ } & g $^{f}$ & r $^{f}$ & \\
VUCD1230+1233    & 2 & 1.0 & 3100 & 12&30&47.40 & 12&33&01.70 &
1390&20 & 19.53 & 18.89  & 9.57 \\ 
VUCD2            & 1 & 1.0 & 7500 & 12&30&48.24 & 12&35&11.10 & 886&19
& 19.21 &18.54  & 11.72 \\ 
VUCD3            & 2 & 1.0 & 3900 & 12&30&57.40 & 12&25&44.80 & 640&11
& 18.87 & 18.07  & 3.00 \\ 
Strom 417        & 1 & 1.0 & 5100 & 12&31&01.29 & 12&19&25.60 & 1872&
15 & 19.69 & 19.01  & 4.97 \\
VUCD1231+1234    & 2 & 1.0 & 4800 & 12&31&02.59 & 12&34&14.11 &
1198&23 & 19.81 & 19.21  & 11.24 \\ 
VUCD5            & 2 & 1.0 & 5400 & 12&31&11.90 & 12&41&01.20 &
1287&13 & 19.10 & 18.39  & 18.39 \\ 
VUCD8            & 1 & 1.0 & 7200 & 12&32&04.33 & 12&20&30.62 &
1679&18 & 19.63 & 19.05  & 18.54 \\ 
VUCD9            & 1 & 1.0 & 6900 & 12&32&14.61 & 12&03&05.40 &
1346&16 & 19.41 & 18.84  & 29.13 \\ 
VUCD1232+0944    & 1 & 0.75 & 6900 & 12&32&55.55 & 09&44&12.70 &
1699&37 & 20.84 & 20.26  & 111.14 \\  
VUCD1233+0952    & 2 & 1.0 & 5100 & 12&33&07.36 & 09&52&54.30 &
1114&18 & 19.97 & 19.39  & 123.27 \\   
\hline
 \multicolumn{15}{l}{$^{a}$Gemini project codes: 1 =
   GN-2007A-Q76, 2 = GN-2008A-Q-92, 3 = GS-2008B-Q-6} \\
 \multicolumn{15}{l}{$^{b}$Total exposure time} \\
 \multicolumn{15}{l}{$^{c}$Projected radius from nearest large galaxy} \\
 \multicolumn{15}{l}{$^{d}$CCD Photometry (Karick in preparation)} \\
 \multicolumn{15}{l}{$^{e}$Photometry by \citet{Cenarro2007}} \\
 \multicolumn{15}{l}{$^{f}$SDSS photometry \citep{Firth2008_paper}}\\
\end{tabular}
\end{table*}

\subsubsection{Gemini-North Virgo Data}
\label{sub-virgo}

The data for Virgo cluster objects were collected for proposals
GN-2007A-Q-76 (P.I.: Firth) and GN-2008A-Q-92 (P.I.: Evstigneeva) in 
February--June 2007 and February--May 2008. We used the long-slit setup
(slit length 5.5~arcmin) of GMOS-N with the B600+\_G5303 grating providing a
wavelength coverage between 3850 and 7050\AA. The slit contains two narrow
bridges splitting it into three equal pieces used to trace and correct the
geometric distortions. The binning of 2$\times$4 yielded a spatial scale of
0.29~arcsec~pix$^{-1}$ and a dispersion of $\approx$1\AA~pix$^{-1}$. The
inter-chip gaps in our setup were around 4500\AA\ and 5760\AA. The slit
width was set to 0.75~arcsec for VUCD1232+0944 and 1~arcsec for
all the remaining targets. Three 2300-sec long exposures were obtained
for every object. Bias, flat and copper-argon arc line frames were
obtained as a part of the standard calibration plan for GMOS-N. Beside
scientific targets, spectra of two spectrophotometric standard stars,
Hiltner~600 and HZ~44 were obtained in order to perform the flux
calibration, as well as the twilight spectral frames which we used to
estimate the spectral line spread of GMOS-N. 

We reduced the data using our own GMOS data reduction pipeline constructed
on top of the universal IFU data reduction toolbox implemented in {\sc idl}.
The data reduction was done independently for every science exposure
(including standard stars and twilight flats).

The data reduction involved several stages. The first was executed for
every individual chip and included the following steps:  
\begin{itemize}
\item Bias subtraction, masking bad columns and hot pixels, removing
cosmic ray hits using the Laplacian filtering technique \citep{vanDokkum2001}.
\item Correcting counts for the read-out analog-to-digital gain.
\item Modelling the global diffuse light by low-order two-dimensional
polynomial surface using counts outside the slit (about 15 binned pixels 
below and above it) and slit bridges and subtracting it.
\end{itemize}

Then, in order to mosaic of individual chips, we used the
transformation coefficients for relative offsets and rotations
determined 
from the IFU arc line and flat frames obtained for a different observing programme
and found in the GEMINI science
archive\footnote{http://www.cadc-ccda.hia-iha.nrc-cnrc.gc.ca/gsa/}.
IFU data frames obtained without the detector binning contain traces of
individual fibers. They allowed us to determine the transformation coefficients
with much higher precision that can be achieved using our binned frames in
the long-slit mode.

We performed the following data reduction steps on the mosaiced frames:
\begin{itemize}
\item The normalised flat field was created by removing the overall
continuum shape by dividing every line by the mean flat field vector along 
the slit; then it was used to correct all science and calibration frames.
\item Arc lines were identified using the line list from the official {\sc
iraf} GMOS reduction pipeline and an iterative procedure was usedto
remove blended lines and those heavily affected by inter-chip gaps and
bad columns on individual detectors. Then, the two-dimensional
dispersion relation was built using the two-dimensional polynomial
surface of the 5rd and 3rd orders along and across the wavelength
range respectively. The RMS of the residuals of the dispersion
relation was usually about 0.09~pix. 
\item We created a model of the night sky airglow spectrum using an approach
proposed by \citet{Kelson2003} aimed at avoiding the artefacts created by the
interpolation of narrow airglow lines. We used the non-linearised science
frame and a two-dimensional dispersion relation in order to construct an
oversampled spectrum of the airglow emission in the wavelength space by
using the slit regions not contaminated by the scientific targets or other
serendipitous targets. This spectrum was then approximated using a smoothing
$b$-spline parametrization with equidistant nodes every 0.3\AA. Then, this
parametrization was used to evaluate the airglow spectrum at every position
on the slit again using the two-dimensional wavelength relation. The sky
subtraction made in this fashion has nearly Poisson quality.
\item After the sky subtraction, we linearised the spectra.
\item Then, we performed the flux calibration using the spectra of
spectrophotometric standard stars.
\end{itemize}

The final step was the optimal extraction of the source on the slit which is
made using an empirically constructed point-spread function from the spectrum
itself as all our targets are sufficiently bright. Then, the spectra from
individual exposures were combined. Since for every target the individual
exposures were obtained during rather short periods of time (an order of a
few days), the individual heliocentric corrections were not required.

The uncertainty frames were computed from the photon statistics and the read-out
noise values, and processed through exactly the same data reduction steps in
order to estimate the flux uncertainties. 

For certain targets, arc line frames were not obtained for every science
exposure, however by comparing the individual science frames, notable shifts
(up-to 2--3 pixels) have been detected both along and across dispersion. For
such situations, we took one of the available arc line frames and shifted it
according to the measured offsets. We stress, that one should not apply
shifts to science exposures as they would introduce additional artefacts
because of the interpolation.

\subsubsection{Gemini-South Fornax Data}
\label{sub-fornax}

The data for Fornax cluster objects were collected for proposal 
GS-2008B-Q-6 (P.I.: Drinkwater) in August--November 2008. As for the Virgo cluster
galaxies, we used the long-slit setup (slit length 5.5~arcmin) of GMOS-S
with a similar B600+\_G5323 grating providing a wavelength coverage between
3900 and 7000\AA. As in GMOS-N, the slit contains two narrow bridges
splitting it into three equal pieces used to trace and correct the geometric
distortions. Similarly, we used binning of 2$\times$4 that yielded a spatial
scale of 0.29~arcsec~pix$^{-1}$ and a dispersion of
$\approx$1\AA~pix$^{-1}$. The slit width was set to 0.75~arcsec for all
targets. An important difference with our GMOS-N observations was the
spectral dithering used to fill the inter-chip gaps: the central wavelength
were different by 50\AA\ for every of the three individual exposures per
target. The GMOS-S calibrations were similar to those obtained with GMOS-N,
only the spectroscopic standard star was different (EG~131).

We reduced the data using the same software as for GMOS-N data with slight
modifications in the two steps.

The arc line spectra obtained during the night time for corresponding
observing blocks were read out only in the central part of the slit of about
1~arcmin in size (truncated arc line frames hereafter). This setup was
sufficient to achieve high-quality wavelength calibration for the science
spectra because all our objects were point sources centered on the slit.
However, the algorithm that we use for the sky subtraction required much
longer slit as it exploited the curvature of spectral lines in the
non-linearised CCD frames. Therefore, we modelled the missing regions of arc
line frames for every observing block using full-frame arc line spectra
obtained at the end of the observing programme.

For our modelling we assumed that between individual observing blocks: (1)
slight overall shifts of the spectra of an order of a few pixels in both
directions may present; (2) slight differences between the wavelength
solutions may present and they could be precisely accounted by the 2nd order
polynomial; (3) there was no significant rotation; (4) the distortion along
the slit remained stable. The relative overall shift across dispersion was
determined by cross-correlating the spectral flat field frames. The relative
position of the slit bridges proved our 4th assumption. The shift along the
dispersion was accounted together with the differences between the wavelength
solutions by individually cross correlating three fragments of the truncated
and full arc line frames, near the centre and at the ends of the spectral
range and then computing the 2nd degree polynomial coefficients describing
the differences. Then, the full-frame arc line spectra were transformed in
order to match these shift using the 2D polynomial image warping and used to
fill the missing data in the truncated arc line frames.

Subsequent tests showed that assumptions 2 (that a 2nd order
polynomial was sufficient) and 3 (that there was no significant
rotation) were justified.  Firstly, we did not detect any measurable
shifts between the truncated arc line frames and the corresponding
parts of the modelled frames. Secondly, we did not have any sky
subtraction artefacts that would be obvious in case of systematic
errors in the wavelength calibration derived from the modelled frames.  

Another difference with the GMOS-N reduction arose from the slightly
different observing strategy: the spectral dithering with a pattern of
50\AA\ was applied in order to fill the inter-chip gaps. However, the
spectrophotometric standard star was observed only in one position.
Therefore, due to the slight change of the grism blaze function as a result
of the central wavelength shift, and because of the slight differences
between the sensitivity curves of individual CCD chips constituting the
GMOS-S mosaic detector, the quantum efficiency curve obtained from the
spectrum of EG~131 cannot be directly applied to calibrate the spectra
obtained with different grism positions. We stress that these effects cannot
be accounted by flat fielding, because the continuum shape of the flat field
lamp was not flat.

In order to perform the flux calibration of the spectra obtained using grism
positions different from that of the EG~131 observation, we considered that
the quantum efficiency curve could be represented as a low-order (9th)
polynomial function with two breaks corresponding to the chip gaps, i.e.
adding two multiplicative coefficients for the 2nd and 3rd CCD chips. We
performed a non-linear minimization in order to determine these coefficients
while the polynomial coefficients were fitted linearly at every evaluation
of the function. The obtained values suggested the differences of an order
of 5--8~per~cent between individual chips that we took into account. This
approach was proved to be valid when we co-added individual spectra for UCDs
since no breaks were detected (either visually or in the residuals of the
full spectral fitting) in the spectral regions around the CCD gaps of
individual observing blocks.

\subsection{Measurement of Spectra}
\label{sub-measurement}

We used two approaches here to estimate the ages and chemical
compositions of the UCDs from the Gemini spectra. Following much of
the previous analysis we started by measuring individual (Lick)
spectral line indices, for comparison to stellar population models. We
also applied a more general approach which derived the parameters by
direct fitting of population models to the whole spectra.  Due to the
relatively low spectral resolution in the selected setups of GMOS-N
and GMOS-S, velocity dispersion measurements were not possible. 

\subsubsection{Lick Index Analysis}
\label{subsub-lick}

In order to measure the Lick indices it was necessary to smooth our spectra to the Lick resolution
of $\sigma_{Lick}\approx 3.6$ \AA,
\citep[see][]{Evstigneeva2007_VUCD}.  This was achieved by convolving the
spectra with Gaussians ($\sigma = 3.03$ \AA\ for the 1 arcsec slit,
and $\sigma = 3.26$ \AA\ for the 0.75 arcsec slit). We then used the {\sc indexf}
program\footnote{
  http://www.ucm.es/info/Astrof/software/indexf/indexf.html}
\citep{Cardiel2007_indexf} 
to measure the Lick/IDS line strength indices, defined
by \citet{Worthey1994_stars}, of each object.  The uncertainties in
the measurements were estimated using the uncertainty arrays described
above, scaled appropriately for the smoothed spectra.
The measured indices, along with the two derived indices $\langle$Fe$\rangle$
and $[$MgFe$]'$ are listed in Table \ref{tab-lick}.
The two derived indices are found as follows: $\langle$Fe$\rangle =$
average of Fe5270 and Fe5335, and $[$MgFe$]'=[$Mg$b (
0.72\times$Fe5270$\ +\ 0.28\times$Fe5335$)]^{1/2}$,
\citep{Thomas2003_SSP}.

\begin{table*}
\caption{Lick/IDS Atomic and Derived Indices}
\label{tab-lick}
\begin{tabular}{lcccccccc}
\hline
Target & \multicolumn{6}{c}{Lick/IDS atomic indices} &
\multicolumn{2}{c}{Derived Indices} \\
 & H$\beta$ & H$\delta$A & H$\gamma$A & Mg $b$ & Fe5270 & Fe5335 &
 $\langle$Fe$\rangle$ & $[$MgFe$]'$ \\
 & (\AA) & (\AA) & (\AA) & (\AA) & (\AA) & (\AA) & (\AA) & (\AA) \\
\hline
FUCD0336-3536 & 2.61 $\pm$ 0.05 & 3.3 $\pm$ 0.1 & 0.91 $\pm$ 0.09 &
0.73 $\pm$ 0.05 & 1.12 $\pm$ 0.05 & 0.59 $\pm$ 0.06 & 0.86 $\pm$
0.04 & 0.84 $\pm$ 0.03 \\
FUCD0336-3522 & 1.57 $\pm$ 0.05 & -0.4 $\pm$ 0.1 & -2.0
$\pm$ 0.1 & 1.83 $\pm$ 0.05 & 2.05 $\pm$ 0.06 & 1.07
$\pm$ 0.07 & 1.56 $\pm$ 0.05 & 1.80 $\pm$ 0.03 \\
FUCD0336-3514 & 1.40 $\pm$ 0.05 & -2.3 $\pm$ 0.14 & -6.4
$\pm$ 0.1 & 4.62 $\pm$ 0.04 & 2.62 $\pm$ 0.05 & 2.17 $\pm$
0.06 & 2.40 $\pm$ 0.04 & 3.39 $\pm$ 0.03 \\
FUCD0336-3548 & 1.99 $\pm$ 0.05 & 1.1 $\pm$ 0.1 & -2.4
$\pm$ 0.1 & 2.31 $\pm$ 0.05 & 1.83 $\pm$ 0.06 & 1.45 $\pm$
0.06 & 1.64 $\pm$ 0.04 & 2.00 $\pm$ 0.03 \\
UCD1          & 1.68 $\pm$ 0.04 & -1.3 $\pm$ 0.1 & -5.31
$\pm$ 0.09 & 3.81 $\pm$ 0.04 & 2.15 $\pm$ 0.05 & 2.13
$\pm$ 0.05 & 2.14 $\pm$ 0.04 & 2.86 $\pm$ 0.03 \\
FUCD0337-3536 & 2.29 $\pm$ 0.05 & -6.6 $\pm$ 0.2 & -2.4
$\pm$ 0.1 & 1.87 $\pm$ 0.05 & 1.92 $\pm$ 0.05 & 1.07
$\pm$ 0.06 & 1.50 $\pm$ 0.04 & 1.77 $\pm$ 0.03  \\
FUCD0337-3515 & 1.76 $\pm$ 0.05 & -0.1 $\pm$ 0.1 & -3.5
$\pm$ 0.1 & 3.50 $\pm$ 0.05 & 2.45 $\pm$ 0.05 & 1.79
$\pm$ 0.06 & 2.12 $\pm$ 0.04 & 2.82 $\pm$ 0.03 \\
FUCD0338-3513 & 2.35 $\pm$ 0.04 & 1.4 $\pm$ 0.1 & -0.96
$\pm$ 0.08 & 1.24 $\pm$ 0.05 & 1.18 $\pm$ 0.05 & 1.09
$\pm$ 0.06 & 1.14 $\pm$ 0.04 & 1.20 $\pm$ 0.03 \\
FUCD0339-3519 & 1.78 $\pm$ 0.05 & -1.5 $\pm$ 0.1 & -5.3
$\pm$ 0.1 & 3.52 $\pm$ 0.05 & 1.97 $\pm$ 0.05 & 1.77
$\pm$ 0.05 & 1.87 $\pm$ 0.04 & 2.60 $\pm$ 0.03 \\
UCD5       & 2.25 $\pm$ 0.05 & 1.4 $\pm$ 0.1 & -1.5
$\pm$ 0.1 & 1.69 $\pm$ 0.05 & 1.40 $\pm$ 0.06 & 1.24
$\pm$ 0.06 & 1.32 $\pm$ 0.04 & 1.51 $\pm$ 0.03 \\
NGC1407GC1 & 2.15 $\pm$ 0.06 & 0.9 $\pm$ 0.2 & -1.7 $\pm$
0.2 & 2.28 $\pm$ 0.06 & 1.95 $\pm$ 0.06 & 1.60 $\pm$
0.07 & 1.78 $\pm$ 0.05 & 2.05 $\pm$ 0.04 \\
VUCD1230+1233 & 1.89 $\pm$ 0.05 & -1.6 $\pm$ 0.1 & -2.8
$\pm$ 0.1 & 2.93 $\pm$ 0.05 & 2.12 $\pm$ 0.05 & 1.81
$\pm$ 0.05 & 1.97 $\pm$ 0.04 & 2.44 $\pm$ 0.03 \\
VUCD2 & 2.21 $\pm$ 0.04 & 0.0 $\pm$ 0.1 & -2.0 $\pm$
0.1 & 1.72 $\pm$ 0.04 & 1.69 $\pm$ 0.04 & 1.28 $\pm$
0.04 & 1.49 $\pm$ 0.03 & 1.65 $\pm$ 0.03 \\
VUCD3 & 1.40 $\pm$ 0.04 & -5.6 $\pm$ 0.1 & -7.6 $\pm$
0.1 & 5.10 $\pm$ 0.03 & 2.55 $\pm$ 0.04 & 2.22 $\pm$
0.04 & 2.39 $\pm$ 0.03 & 3.54 $\pm$ 0.02 \\
Strom 417 & 2.04 $\pm$ 0.04 & -0.9 $\pm$ 0.1 & -1.2 $\pm$
0.1 & 2.82 $\pm$ 0.04 & 2.31 $\pm$ 0.04 & 1.88 $\pm$
0.04 & 2.10 $\pm$ 0.03 & 2.48 $\pm$ 0.02 \\
VUCD1231+1234 & 1.98 $\pm$ 0.06 & 0.8 $\pm$ 0.1 & -1.2
$\pm$ 0.1 & 1.29 $\pm$ 0.06 & 1.23 $\pm$ 0.06 & 0.51
$\pm$ 0.07 & 0.87 $\pm$ 0.05 & 1.15 $\pm$ 0.04 \\
VUCD5 & 1.73 $\pm$ 0.03 & -1.69 $\pm$ 0.09 & -4.71 $\pm$
0.07 & 3.34 $\pm$ 0.03 & 2.21 $\pm$ 0.03 & 1.81 $\pm$
0.04 & 2.01 $\pm$ 0.03 & 1.15 $\pm$ 0.04 \\
VUCD8 & 2.13 $\pm$ 0.04 & 0.6 $\pm$ 0.1 & -2.22 $\pm$
0.09 & 1.47 $\pm$ 0.04 & 1.65 $\pm$ 0.04 & 1.10 $\pm$
0.05 & 1.38 $\pm$ 0.03 & 1.48 $\pm$ 0.03 \\
VUCD9 & 2.43 $\pm$ 0.04 & -0.17 $\pm$ 0.09 & -2.61 $\pm$
0.08 & 2.07 $\pm$ 0.03 & 1.77 $\pm$ 0.04 & 1.60 $\pm$
0.04 & 1.69 $\pm$ 0.03 & 1.89 $\pm$ 0.02 \\
VUCD1232+0944  & 1.3 $\pm$ 0.1 & -6.5 $\pm$ 0.5 & -4.3 $\pm$
0.3 & 1.27 $\pm$ 0.09 & 2.01 $\pm$ 0.08 & 1.5 $\pm$
0.1 & 1.76 $\pm$ 0.06 & 1.54 $\pm$ 0.06 \\  
VUCD1233+0952  & 2.12 $\pm$ 0.07 & -2.0 $\pm$ 0.2 & -2.6 $\pm$
0.2 & 2.02 $\pm$ 0.07 & 2.16 $\pm$ 0.07 & 1.19 $\pm$
0.08 & 1.68 $\pm$ 0.05 & 1.95 $\pm$ 0.04 \\  
\hline
\end{tabular}
\end{table*}

The Lick indices offer the advantage of a substantial body of previous
measurements for comparison and discussion, but they do not make use
of all the information available in the spectral data. One way to
improve the analysis is to combine the values of multiple Lick indices
to obtain more accurate results, as done by
\citet{Proctor2004_indices}. In Section~\ref{sub-spectral} we take a
slightly different approach, by fitting models directly to the full
spectra themselves rather than first calculating line indices. 

\subsubsection{Full spectral fitting}
\label{sub-spectral}

We used the {\sc nbursts} full spectral fitting technique
\citep{Chilingarian2007a,Chilingarian2007b} with the {\sc pegase.hr}
\citep{LeBorgne2004} simple stellar population (SSP) models in order
to determine SSP-equivalent ages and metallicities.
 
The {\sc nbursts} full spectral fitting package implements a pixel-space
fitting algorithm. Generally, an observed spectrum is approximated by a
linear combination of stellar population models broadened with the galaxy's
parametric line-of-sight velocity distribution, whose parameters (e.g. age,
metallicity, initial mass function) are determined inside the same
minimization loop as the internal kinematics. The fitting procedure includes
a multiplicative polynomial continuum aimed at absorbing possible flux
calibration issues both, in observations and in the models. The {\sc
nbursts} technique is shown to produce unbiased esimates of ages and
metallicities for $\alpha$-enhanced stellar populations
\citep{Chilingarian2008abell} although the {\sc pegase.hr} stellar population
models are not representative of such populations because they are
constructed from empirical stellar spectra for stars mostly from the Solar
neighbourhood which are known to have their $\alpha$/Fe abundance ratios
anti-correlated with metallicities.

In this study we use the simplest case of the fitting procedure, when the
observed spectrum is approximated by one SSP model characterised by its age
and metallicity. We used an SSP grid computed with the {\sc pegase.hr}
evolutionary synthesis code at the intermediate spectral resolution
(R=10000) in a waveleneth range 3900--6800~\AA\ for the \citet{Kroupa1993}
stellar initial mass function. We used the 29th order multiplicative
polynomial continuum. The model SSP grid was pre-convolved with spectral
line spread functions of GMOS-N and GMOS-S spectrographs varying with the
wavelength. We computed them by fitting the twilight spectra obtained in the
same instrumental setups as UCD spectra against the R=10000 Solar spectrum
from the {\sc elodie.3} \citep{Prugniel2004} stellar library in five wavelength
segments between $3900 < \lambda < 6800$~\AA\ using the {ppxf} procedure
\citep{Cappellari2004}. For the 1-arcsec wide slit of GMOS-N, the spectral line spread
changes from $\sigma_{\rm{inst}} = 150$~km~s$^{-1}$ at 3900~\AA\ to
100~km~s$^{-1}$ at 6700~\AA, with nearly constant Gauss-Hermite coefficients
$h3$ and $h4$ \citep{vanderMarel1993} of about 0.00 and $-0.10$ correspondingly. For
the 0.75-arcsec wide slit of GMOS-S, the spectral resolution was slightly
better and corresponded to $\sigma_{\rm{inst}} = 120$~km~s$^{-1}$ at
3900~\AA\ and 70~km~s$^{-1}$ at 6700~\AA\ with the $h3$ and $h4$ values of
0.00 and $-0.08$.

In addition to the best-fitting values of radial velocity and SSP-equivalent
age and metallicity (included in Table~\ref{tab-age-metallicity}) which are
obtained in a single minimization loop, for every object we computed a
confidence map in the age--metallicity space similar to those
presented in \citep{Chilingarian2008fornax,Chilingarian2011}.  These can be
found in Appendix~\ref{ap-fig}. We notice, that when the
{\sc NBursts} technique is used in the $\chi^2$-mapping mode, i.e. fitting a
single SSP with fixed age and metallicity at every grid node, the algorithm
``degenerates'' into the form totally equivalent to {\sc ppxf} \citep{Cappellari2004}
with a single template spectrum.

\section{Analysis}
\label{sec-analysis}

In this section we measure the internal composition of the UCDs from
the Gemini spectra. We first compare the values of individual line
indices to models before directly fitting stellar population models to
the spectra. 

\subsection{$\alpha$-element Abundance Ratios from Lick Indices}
\label{sub-alpha}

The star formation history of stellar systems, such as dwarf galaxy
nuclei and GCs, can be probed with $\alpha$-element abundance ratios
\citep[e.g.][respectively]{Geha2003_dEN,Brodie2006_GCs}.   

In simple closed models of stellar evolution
\citep[e.g.][]{McWilliam1997_alpha, Bressan1994}, the average
metallicity ([Fe/H]) of the populations always increases with time as
metals are formed in stars and ejected into the ISM by supernovae. The
high-mass supernovae (type II) produce large amounts of the
``$\alpha$-elements'' (such as oxygen and magnesium) by neutron
capture processes leading to a high $\alpha$-element abundance
([$\alpha$/Fe]). After a delay (due to their lifetimes) the low-mass
stars stars will start producing type Ia supernovae: these mainly
produce iron, so the alpha abundance then starts to decrease. 

In a given observed stellar population, low (Solar or sub-Solar)
$\alpha$-abundances reflect star formation which has taken place at a
low rate over a long period, so that the type Ia supernovae have
significantly reduced the [$\alpha$/Fe] ratio. The cores of dwarf
galaxies show these properties \citep{Geha2003_dEN,
    Gorgas1997}. In contrast, high (super-Solar) $\alpha$-element
abundances reflect a rapid early stage of star formation which formed
most of the population. This produces much larger amounts of the
$\alpha$-elements before the SN Ia start, so the [$\alpha$/Fe] ratio
remains high. Globular clusters typically have super-Solar abundances
\citep{Cohen1998_GCs} having formed all their stars very early in the
Universe.

\begin{figure}
\epsfig{file=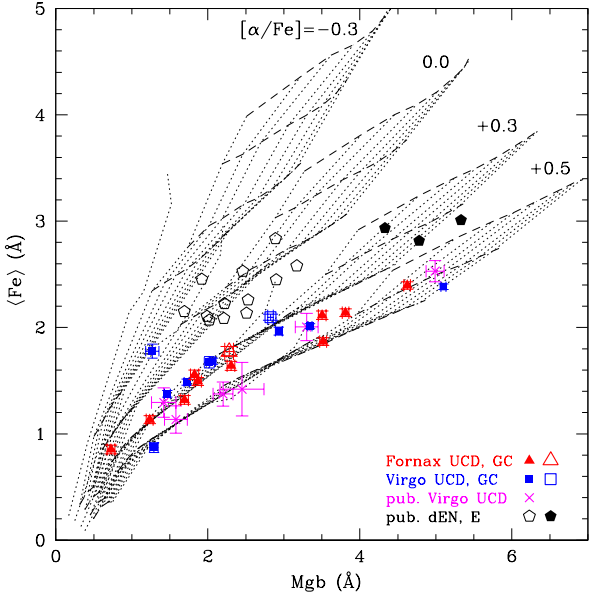,width=1.0\linewidth,angle=0}
\caption{
  $\alpha$-element abundances of UCDs based on Lick index
  measurements. SSP model grids are plotted for a range of
  [$\alpha$/Fe] values from \citet{Thomas2011SSP}. The grid lines
  simulate ranges of constant age and metallicity.  The Gemini data
  are plotted as (red) triangles for the Fornax
    objects and (blue) squares for the Virgo
    objects. In each case the solid symbols are UCDs and the open
  symbol is a GC. Previous measurements of Virgo UCDs
  \citep{Evstigneeva2007_VUCD} are plotted as (magenta) crosses.
  Previously published data are shown for comparison: nucleated dE
  galaxies \citep[unfilled pentagons][]{Geha2003_dEN} and elliptical
  galaxies \citep[filled pentagons][]{Trager2000_E}. 
}
\label{fig-MgbFe}
\end{figure}

We present measurements of the Lick indices sensitive to
$\alpha$-element abundance for the UCDs in Figure \ref{fig-MgbFe}. The
figure plots the derived iron index $\langle$Fe$\rangle$ (see previous
section) against Mgb. The figure also shows predicted model grids for
populations with a range of abundance
([$\alpha$/Fe]$=-0.3,0,0.3,0.5$). The models were calculated by
\citet{Thomas2011SSP}. These new models have been calculated for Lick
indices measured with flux-calibrated spectra, so do not require any
zero-point corrections \citep[such as used in our earlier analysis
---][]{Firth2009_UCD}. 

The figure shows that the majority of the UCDs in both clusters have
super-Solar $\alpha$-element abundances, mostly consistent with the
[$\alpha$/Fe]$=0.3$ model. However, some of the
  UCDs with higher $\langle$Fe$\rangle$ values are also consistent
  with the [$\alpha$/Fe]$=0.5$ model. These high abundances are
consistent with both old elliptical galaxies (shown as crosses on the figure) and
globular clusters \citep[open triangle and square, but also see Fig.~2
of ][]{Thomas2011SSP}. The average abundance of the UCD population
(parameterised by the ratio $\langle$Fe$\rangle /$Mgb)  is
significantly higher ($p=0.003$ using a T-test) than that of nucleated
dwarf galaxies (shown as open pentagons in the figure). There is no
significant difference between the abundances of the Virgo and Fornax
UCDs. We note that one UCD (VUCD1232+0944) has a lower abundance,
consistent with the Solar abundances of the dE,N galaxies.  


\subsection{Metallicity and Age from Lick Indices}
\label{sub-age}


Several of the Lick indices are sensitive to metallicity and age. We chose to focus on the combination of H$\beta$ and the compound index $[$MgFe$]'$ for ease of comparison with previous work \citep[e.g.][]{Firth2009_UCD}. In addition, the $[$MgFe$]'$ index has the advantage that it measures metallicity without being affected by $\alpha$-element abundance \citep{Thomas2003_SSP}.

\begin{figure}
\epsfig{file=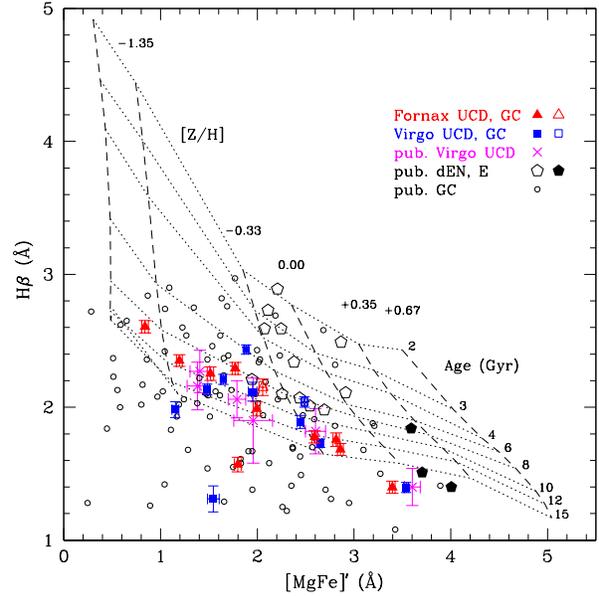,width=1.0\linewidth,angle=0}
\caption{Age and metallicity of UCDs as indicated by the H$\beta$ and $[$MgFe$]'$ Lick indices.  SSP model grids are plotted for [$\alpha$/Fe]$=0.3$ dex from \citet{Thomas2011SSP}. The grid lines are simulated constant age and metallicity. The symbols are as in the previous figures with the addition of M87 (Virgo Cluster) globular clusters \citep[circles][]{Cohen1998_GCs}.}
\label{fig-HbMgFe}
\end{figure}

\begin{figure}
\epsfig{file=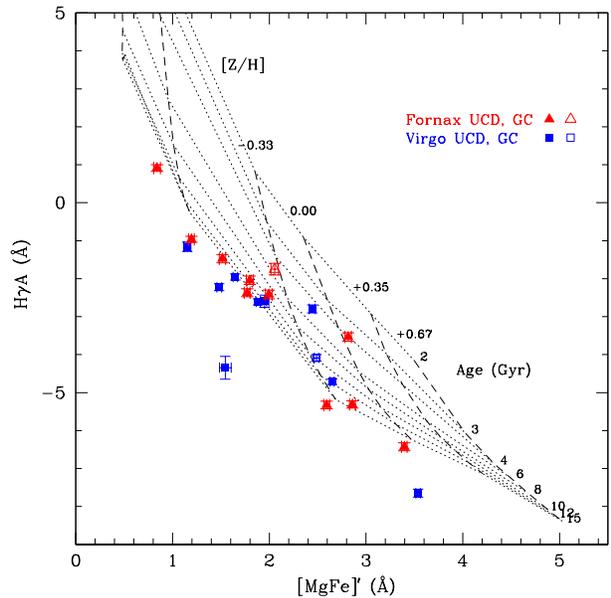,width=1.0\linewidth,angle=0}
\caption{Age and metallicity of UCDs as indicated by the H$\gamma$ and $[$MgFe$]'$ Lick indices.  SSP model grids are plotted for [$\alpha$/Fe]$=0.3$ dex from \citet{Thomas2011SSP}. The grid lines are simulated constant age and metallicity. The comparison objects from the previous plots are not shown. This plot shows that, although the H$\gamma$ index does not display as much scatter as H$\beta$, the UCD have similar properties compared to the SSP model.
}
\label{fig-HgMgFe}
\end{figure}

We show these indices measured for the UCDs in Figure~\ref{fig-HbMgFe}, along with the predicted model grid for a SSP model \citep{Thomas2011SSP} with the super-Solar abundance ($[\alpha$/Fe$]\simeq0.3$) found in the previous section for these objects. The figure also shows for comparison purposes the positions of samples of giant elliptical galaxies, nucleated dwarf elliptical galaxies, and globular clusters. The dE,N galaxies display the youngest ages, with slightly sub-Solar metallicities. The giant ellipticals are older, with higher metallicities, whereas the globular clusters are also old, but with low metallicities \citep[see][]{Cohen1998_GCs, Geha2003_dEN}. We note that the globular clusters in Figure~\ref{fig-HbMgFe} show a considerable scatter {\em below} the lowest (maximum age) of the model grid lines. This is a known effect, possibly related to the modelling of hot giant stars \citep{Thomas2011SSP}. The higher order H$\gamma$ index shows less scatter, as shown in Figure~\ref{fig-HgMgFe} (without the comparison objects), but the overall conclusions are not changed (i.e.\ the UCD positions compared to the models).


We inverted the SSP models \citep{Thomas2011SSP} plotted in Figure~\ref{fig-HbMgFe} to obtain estimates of the ages and metallicities of each object using a binomial polynomial interpolation \citep{Blazquez2006,Cardiel2003,Wolberg1990}.  The corresponding ages and metallicities obtained from the Lick indices are given in Table~\ref{tab-age-metallicity}.


Both of the figures show the UCDs clustering along the oldest model track, but with a considerable range of metallicity. There is a scatter below the oldest model track in the H$\beta$ plot but this is consistent with the scatter observed for the globular clusters. This scatter in age leads to some overlap with the dE galaxies, but the mean age of the UCD sample (as parameterised by $[$MgFe$]'/2 + $H$\beta$)  is significantly ($p<0.001$ with a T-test) older than that of the dE population. As above, we find no evidence for any significant age difference between the Virgo and Fornax UCDs. The UCDs have a much wider range of metallicity than any of the comparison galaxy populations: the range is however consistent with that of the comparison GC population \citep[from M87 in the Virgo Cluster; ][]{Cohen1998_GCs}. The UCD population therefore includes a metal-rich component like many GC populations.

We also show some previous UCD measurements using the Keck Telescope \citep{Evstigneeva2007_VUCD} in Figures \ref{fig-MgbFe} and \ref{fig-HbMgFe}. The object with the highest metallicity from the Keck measurements is the same object we have repeated in our Gemini observations, VUCD3. In each case both measurements are consistent, giving an indication of the repeatability of our measurements.


One object, VUCD1232+0944 in the Virgo Cluster, has extreme values in all the index plots. It has the lowest abundance of the UCDs (Solar) in Figure~\ref{fig-MgbFe} and also the lowest H$\beta$ index in Figure~\ref{fig-HbMgFe}. The H-gamma value also shows this object lying below the main population. We inspected the spectrum of this object and did not find any unusual features. The population model gives an age of 12.8 Gyr and $[Fe/H] = -1.0$, which is consistent with the Lick results given the large scatter in the H$\beta$ index.

\begin{table*}
\centering
\caption{Age and Metallicity results from Lick indices and Full
  Spectral Fitting}
\label{tab-age-metallicity}
\begin{tabular}{l r@{ $\pm$ }l r@{ $\pm$ }l r@{ $\pm$ }l r@{ $\pm$ }l }
\hline
 & \multicolumn{4}{c}{from Lick Indices:} & \multicolumn{4}{c}{from
   Spectral Fitting: $^{a}$} \\
 Target & \multicolumn{2}{c}{Age} & \multicolumn{2}{c}{Metallicity} &
 \multicolumn{2}{c}{Age}  & \multicolumn{2}{c}{Metallicity} \\
 & \multicolumn{2}{c}{(Myr)} & \multicolumn{2}{c}{(dex)} &
 \multicolumn{2}{c}{(Myr)} &  \multicolumn{2}{c}{(dex)} \\
\hline
FUCD 0336-3536 & 7\,980&200 & -1.58&0.04 & \multicolumn{2}{c}{
  $>15\,000\,^{b}$ } & -1.76&0.02 \\ 
 FUCD0336-3522 & 11\,030&160 & -0.70&0.01 & 8\,180&230 & -0.82&0.02 \\ 
FUCD0336-3514 & 19\,280&700 & -0.04&0.02 & \multicolumn{2}{c}{$>15\,000$}
& -0.23&0.01 \\
FUCD0336-3548 & 7\,670&250 & -0.55&0.02 & 7\,980&210 & -0.70&0.02 \\
UCD1          & 9\,710&260 & -0.06&0.01 & 13\,760&350 & -0.42&0.01 \\
FUCD0337-3536 & 5\,400&170 & -0.60&0.02 & 7\,350&150 & -0.87&0.02 \\
FUCD0337-3515 & 8\,410&260 & -0.06&0.02 & 10\,750&250 & -0.49&0.02 \\
FUCD0338-3513 & 9\,530&580 & -1.20&0.03 & 6\,236&99 & -1.16&0.02 \\
FUCD0339-3519 & 13\,000&1\,100 & -0.32&0.02 & 12\,980&340 & -0.46&0.01 \\
UCD5          & 12\,450&280 & -0.97&0.03 & 7\,210&160 & -1.01&0.02 \\
NGC1407GC1    & 6\,760&340 & -0.47&0.03 & 6\,290&200 & -0.67&0.02 \\
VUCD1230+1233 & 11\,380&830 & -0.33&0.02 & 9\,730&230 & -0.76&0.02 \\

VUCD2         & 5\,473&87 & -0.49&0.01 & 9\,950&200 & -0.98&0.01 \\
VUCD3         & 24\,600&2\,300 & -0.28&0.09 &
\multicolumn{2}{c}{$>15\,000$} & -0.15&0.01 \\
Strom417 & 4\,090&170 & -0.05&0.01 & 8\,500&150 & -0.61&0.01 \\
VUCD1231+1234 & 14\,080&110 & -1.33&0.03 & \multicolumn{2}{c}{
  $>15\,000\,^{b}$ } &  -1.67&0.02 \\
VUCD5 & 14\,140&690 & -0.30&0.01 & 12\,240&220 & -0.57&0.01 \\
VUCD8 & 12\,610&630 & -1.03&0.02 & 8\,550&160 & -1.08&0.01 \\
VUCD9 & 4\,520&130 & -0.41&0.01 & 6\,812&98 & -0.87&0.01 \\
VUCD1232+0944  & 55\,300&8\,800 & -0.83&0.05 & 12\,830&800 &
-1.02&0.02 \\ 
VUCD1233+0952  & 6\,260&270 & -0.37&0.02 & 10\,910&450 & -0.99&0.02 \\
\hline
\multicolumn{9}{l}{$^{a}$ Age-metallicity confidence maps for these
  results can be found in Appendix~\ref{ap-fig}} \\
\multicolumn{9}{l}{$^{b}$ The spectral fitting reached the upper age
  limit of the model grid.} 
\end{tabular}
\end{table*}

\subsection{Metallicity and age from the full spectral fitting}

\begin{figure}
\centering
\epsfig{file=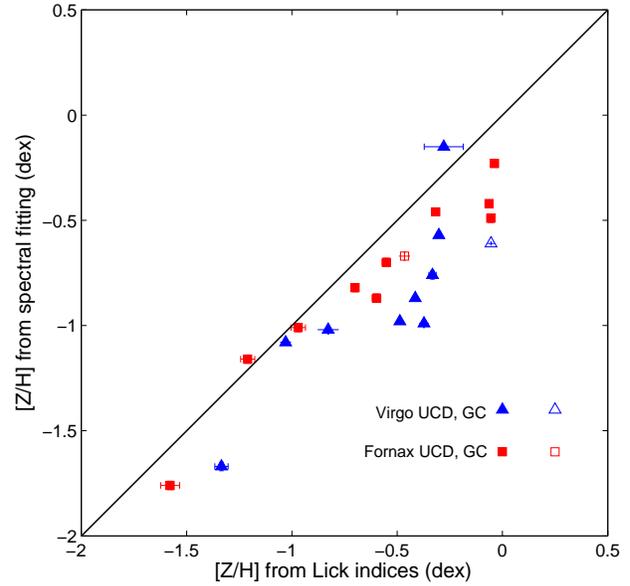,width=1.0\linewidth,angle=0,bb=
 80   202   510   620, clip=true}
\caption{
  Comparison of metallicities calculated from the
  Lick indices and from the full spectral fitting. The theoretical line
  corresponding to full agreement between the methods is indicated in
  black: there is an offset of about 0.2 dex as discussed in the text.}
\label{fig-comparison-metallicity}
\end{figure}

The {\sc NBursts} spectral fitting approach makes full use of the
available wavelength range and spectral resolution of the data so it
should provide more reliable ages and metallicities than those derived
from the line indices which only use a small fraction of the
available age- and metallicity-sensitive data contained in the index
definition regions. However, a comparison of the metallicity estimates
in Figure~\ref{fig-comparison-metallicity} reveals a systematic
metallicity-dependent offset: the {\sc NBursts} metallicities are
about 0.2 dex lower than the Lick values at high metallicity.


This offset may be caused by the models used by {\sc NBursts}: the age
estimates from the full spectral fitting are not affected by non-solar
abundance ratios of stellar populations \citep[see
e.g.][]{Chilingarian2008abell} but the metallicities can be affected. This
is a result of the properties of the {\sc pegase.hr} stellar
population models used by {\sc NBursts}.  These models are based on an
empirical stellar library containing only stars in the Solar vicinity
(see above) having a characteristic abundance pattern, representative
of $\alpha$-enhanced ($[$Mg/Fe$] \approx +0.3$~dex) populations at low
iron metallicities ($[$Fe/H$]\lesssim -1.0$~dex) but Solar abundances
($[$Mg/Fe$] \approx 0.0$~dex) at high, i.e.  Solar and slightly
sub-Solar, iron abundances. Most of the objects in our study are
$\alpha$-enhanced (see above).  Although the spectral region contains
many features of $\alpha$-elements between 4200\AA\ and 5200\AA, the
fitting residuals of metal-rich UCDs such as VUCD3 exhibit a strong
template mismatch at the Mg$b$ triplet position. Therefore the fitting
procedure is representative of iron metallicities when fitting spectra
$\alpha$-enhanced metal-rich objects where models are Solar-scaled in
[$\alpha$/Fe]. At the same time, for metal-poor targets we are
measuring the mean because the models themselves are representative of
$\alpha$-enhanced populations. This explains why at high metallicities
our $[$Z/H$]$ are slightly underestimated (about 0.2~dex) compared to
the abundance-insensitive measurements done with the Lick indices.

Generally, ages are determined with much higher
  uncertainties (in dex) than for the metallicities.  This illustrates
that most features in the spectrum are much less sensitive to age
than to metallicity.  For this reason we have not made a detailed
comparison of the ages derived by the two methods.

Confidence level maps of stellar population parameters in the
age--metallicity space demonstrate quite well a well-known age--metallicity
degeneracy which is manifested by the inclined shape of error contours so
that older ages may ``compensate'' lower metallicities. The degeneracy is
much weaker though for metal-poor objects. Generally, ages are determined
with a much higher uncertainties (in dex) than metallicities that
illustrates the fact that most features in the spectrum are much less
sensitive to the age than to the metallicity.

In many cases (5 in Fornax and 5 in Virgo) we observe double minima in the
age--metallicity confidence level maps with one of them always at very old
age values significantly exceeding the presently accepted age of the
Universe (13.7~Gyr) and reaching the maximal age in the model grid (17~Gyr).
Their origin is unclear and it is probably connected to the model
imperfections in the selected wavelength range. Although in a few cases
those ``old'' minima are primary (i.e. $\chi^2$ has a global minimum there), we
discard all age estimates exceeding 15~Gyr and consider the secondary
minimum if it exists.

Four of the objects from our present Fornax Cluster sample have
been previously analysed using the {\sc NBursts} technique applied to
high-resolution spectra: FUCD0337-3538$=$UCD1, FUCD0339-3504$=$UCD5,
FUCD0336-3514$=$ucd329.7 \citep{Chilingarian2008abell} and FUCD0339-3519$=$F-23
\citep{Chilingarian2011}. For three of them our age and metallicity
estimates agree very well with the published data but have much better
precision. For UCD5, our present estimate falls within the published
2-$\sigma$ uncertainty level which can be explained by the low
metallicity of this object making it a subject to a strong
age--metallicity degeneracy and hence hampering the determination of
its stellar population properties in the short wavelength range of
FLAMES/Giraffe data analysed by \citet{Chilingarian2008abell}.


\section{Discussion}
\label{sub-theories} 

In this section we discuss the implications of the age and metallicity
measurement presented above in terms of formation theories for
UCDs. Both the Lick index analysis and the direct model fitting reveal
the UCDs to be an old population with super-Solar abundances and a
wide range of metallicity. More generally, we also compare the
properties of the UCDs to a wide range of other stellar
systems. Almost without exception they have very high metallicities
for their luminosities. 

Starting with the Lick index analysis, we found that the UCDs in both
clusters had generally high (super-Solar) alpha element abundances,
implying short formation times \citep{Matteucci1994,Thomas2005}.
Similar values are observed in most globular cluster populations, as
distinct from present-day galaxies (both dwarf and giant) which have
lower abundances, indicative of some continuing star
formation. At first sight, this observation could be taken as evidence
against the threshing model in that the UCDs have higher abundances
than dwarf elliptical galaxies. However, as noted by
\citet{Evstigneeva2008}, if the UCDs were stripped from dwarf galaxies
at an early epoch, and their gas was removed at the same time, this
would prevent further star formation, also leading to high
abundances. 

The generally old ages of the UCDs also distinguish them from
present-day (nucleated) dwarf elliptical galaxies. This too argues
against recent disruption, but not against an early disruption. The
spread of age and metallicity of the UCDs is consistent with that
observed for globular clusters, notably those in the Virgo
Cluster. These results are generally consistent with earlier work, but
our increased sample has revealed a larger range of metallicity. We
find metallicities as low as $-1.6$ dex (one in each cluster:
VUCD1231+1234 and FUCD0336-3536) according to the model fits. This is
in contrast to \citet{Chilingarian2008fornax} who measured a smaller UCD
sample, finding them all to be metal-rich ([Fe/H]$>-1$), although the
larger sample measured by \citet{Chilingarian2011} included three
objects with lower metalicity ($-1.39<$[Fe/H]$<-1.2$). We are now
finding that UCDs exhibit the full range of (low) metalicities shown
by globular cluster populations. 

Our data also allow us to compare the UCD populations between the
Virgo and Fornax clusters as we used very similar Gemini (North and
South) data for both clusters and identical analysis. Furthermore the
objects observed had the same luminosity distribution in both
clusters. Using the model fits we obtain mean ages of $(10.4 \pm 1.1)$
Gyr for Fornax and  $(11.2 \pm 0.9)$ Gyr for Virgo with no significant
differences in the distributions\footnote{The Kolmogorov-Smirnov,
  t-test and F-test all gave $p>0.5$ that the samples were from the
  same distributions.}. Similarly there is no significant difference
in the mean metalicities of [Fe/H] $=-0.79 \pm 0.14$ for Fornax and
[Fe/H] $=-0.90 \pm 0.14$ for Virgo. (In each case,
  for age and metallicity we quote the standard error of the
  respective mean value.) This contradicts the earlier
suggestions made \citep{Mieske2006_UCD, Firth2009_UCD} that the Fornax
UCD population was younger than that of Virgo due to differences in
the cluster environment. The cluster environments are different (Virgo
is larger and much less relaxed), but UCDs are mostly only found close
to the central galaxies of each cluster, so we would argue that they
experience similar local environments in the two clusters, resulting
in similar internal properties. 

For two of the Virgo cluster objects with HST imaging, VUCD~3 and
VUCD~5, \citet{Evstigneeva2007_VUCD} published dynamical masses and
$V$-band dynamical mass-to-light ratios: $(M/L)_{\rm{dyn}}=5.4 \pm
0.9$ and $3.9 \pm 0.6$ in Solar units. Using our precise age and
metallicity measurements, we can derive their stellar masses and hence
estimate their dark matter fractions defined as
$((M/L)_{\rm{dyn}}-(M/L)_{*})/(M/L)_{\rm{dyn}}$. The $V$-band stellar
mass-to-light ratios for VUCD~3 and VUCD~5 are $(M/L)_{*}=4.5 \pm 0.2$
($8.1 \pm 0.4$) and $2.85 \pm 0.13$ ($5.0 \pm 0.2$) $(M/L)_{\odot, V}$
respectively, where values in parentheses correspond to the
\citet{Salpeter1955} IMF. The corresponding dark
mass fractions are: $15 \pm 25$ ($-50 \pm 25$) and $25 \pm 20$
($-28 \pm 20$) per cent.  Negative values indicate that the stellar
mass estimate exceeds the dynamical one, suggesting that the Salpeter
IMF is not compatible with the observations \citep[see discussion
by][]{Chilingarian2011}.

The fact that we do not detect dark matter in these two Virgo UCDs is
in contrast to previous work suggesting that the Virgo UCDs had higher
mass-to-light ratios than the Fornax UCDs. \citet{Hasegan2005}
reported mass-to-light ratios between 6 and 9 for their ``probable''
UCDs, significantly higher than comparison measurements for UCDs in
the Fornax Cluster (1-3; see their Figure 11). Although we do not have
data for a large sample of Virgo UCDs, VUCD~3 and VUCD~5 are the most
massive UCDs in Virgo and for these at least we find no evidence of
dark matter.


\begin{figure*}
\centering
\epsfig{file=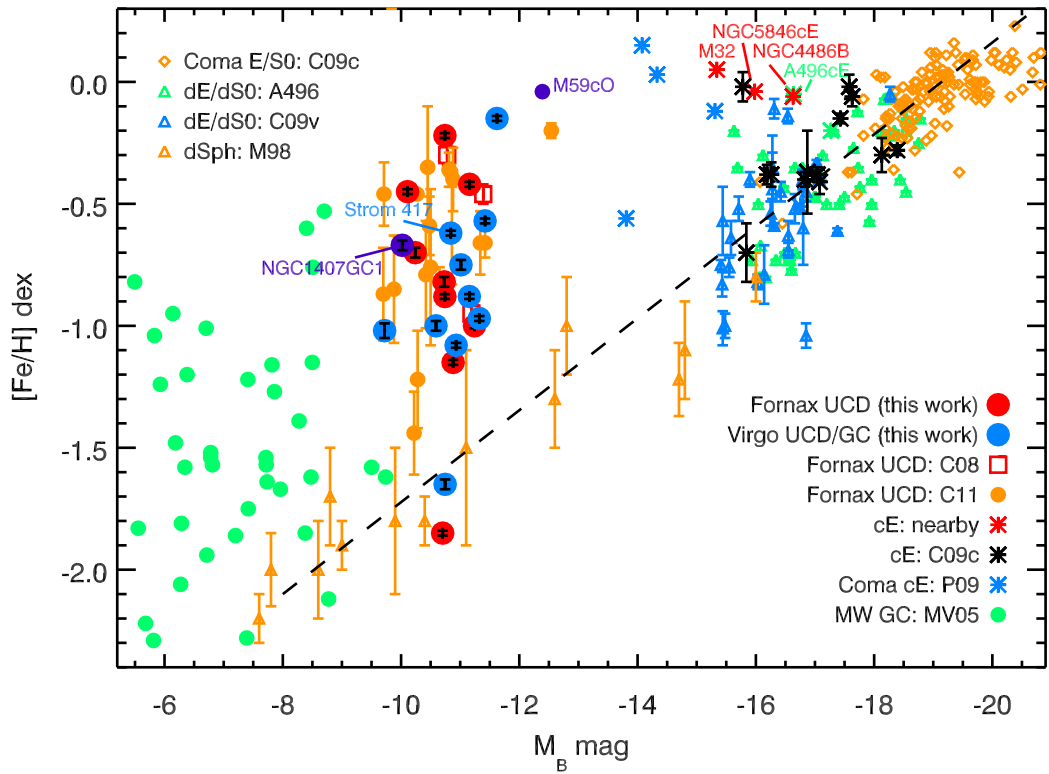,width=0.8\linewidth,angle=0}
\caption{
  The metallicity-luminosity relation for UCDs and other stellar systems. The model-fitted
  metalicities of each UCD are plotted against their absolute $B$
  magnitudes. The dashed line traces the canonical
  metallicity-luminosity relation for early-type galaxies \citep[e.g.\ Fig.~8 of][]{Chilingarian2011}. Comparison values for different types of galaxy and globular clusters are taken from the literature according to the following codes:
A496 - \citet{Chilingarian2008abell}
C08    - \citet{Chilingarian2008fornax},
C09c  - \citet{Chilingarian2009cE},
C09v - \citet{Chilingarian2009virgo}, 
C11 - \citet{Chilingarian2011},
M98  - \citet{Mateo1998}, 
MV05 -  \citet{McLaughlin2005}, 
P09  - \citet{Price2009}.
Data for the nearby cE galaxies are from
\citet{Chilingarian2008m59} (M59cO),
\citet{Chilingarian2010n5846} (N5846cE),
\citet{SanchezBlazquez2006} (N4486B),
\citet{Graham2002} (M32 luminosity) and \citet{Worthey2004} (M32
metallicity).
} 
\label{fig-lmetal}
\end{figure*}

%

In Fig.~\ref{fig-lmetal} we plot the model-fitted UCD metallicities
against their luminosities compared to previous measurements of  
UCDs and dwarf galaxies, as well as reference data for globular
clusters and elliptical galaxies.  (We use the
  model metallicities rather than the those derived from the Lick
  indices as they use more of the spectra dalta and have smaller
  statistical uncertainties.) It is immediately obvious that our
data are much more precise than most of the previous work. With this
precision, the two UCDs with lower metallicity stand out as being
significantly different to the rest of the UCD population. We should
note, however, that the nature of our sample selection means that we
cannot exclude future measurements of intermediate objects. In fact,
with increasing measurements we now see there is a continuous range of
objects with intermediate properties (both luminosity or metallicity)
filling the space between UCDs and compact elliptical (``cE'' or
``M32-like'') galaxies. 

The UCDs in Fig.~\ref{fig-lmetal} with new Gemini measurements show no
evidence of a metallicity-luminosity relation. We also measured the
metallicity of the objects as a function of projected distance from
the nearest large galaxy in both clusters and found no correlation in
those parameters. In particular the two objects from the intra-cluster
Virgo field (VUCD1232+0944 and VUCD1233+0952 at 550 kpc from M87) have
metallicities entirely consistent with the other Virgo objects. 

Fig.~\ref{fig-lmetal} does, however, show that (with two exceptions)
the UCDs lie well above the canonical metallicity-luminosity trend for
early-type galaxies (including nucleated dwarf elliptical galaxies),
as was demonstrated previously for Fornax Cluster UCDs
\citep{Chilingarian2011}. This is exactly what we might expect from
tidal disruption of current-day dwarf galaxies: the luminosity would
decrease while the metallicity would remain high. On the other hand,
the figure also shows that many globular clusters (albeit at lower
metalicities and luminosities) also lie above the canonical
metallicity-luminosity trend. Thus the UCDs could represent either the
high-luminosity end of the globular cluster metallicity-luminosity
relation, or the natural end point of a constant-metallicity stripping
process of dwarf galaxies. There are now sufficient intermediate
objects on both sides that the data shown in this figure cannot rule
out either hypothesis. A promising approach to help separate the two
populations would be to measure the physical sizes of the
objects. This has recently been done by \citet{Brodie2011_new} who found
evidence that UCDs in the Virgo Cluster formed a distinct population
with significantly larger sizes than globular clusters.

\section{Summary}
\label{sec-summary}
For each of our Fornax cluster and Virgo cluster UCDs, we measured
both the Lick/IDS line strength indices and performed the {\sc
  nbursts} full spectral fitting technique, and compared these results
to SSP models.  Our findings are summarised below:

\begin{itemize} \itemsep12pt \parskip0pt \parsep0pt

  \item While much of the evidence, such as super-Solar
    $\alpha$-abundance ratios, old age measurements, and a large
    spread in the metallicity measurements,  seems to imply that the
    formation of UCDs is consistent with that of GCs, it does not rule out
    the threshing formation model.  The data are only inconsistent
    with stripping of {\em present-day}  nucleated dwarf galaxies, but
    not with much older disruptions before the nuclei evolved to their
    current composition.

  \item Many of our measurements contrast those from previous work.
    The differences are outlined below:
    \begin{enumerate} \itemsep6pt \parskip0pt \parsep0pt
      \item Our metallicity measurements exhibit a larger range of
        metallicity than indicated in previous work, consistent with
        the full range of metallicities of GCs.
      \item Our age and metallicity measurements, along with
        luminosity distributions, indicate similar UCD populations in
        both the Virgo and Fornax clusters.  Previous work suggests
        that the different cluster environments cause the Fornax UCD
        population to be younger than that of Virgo.
      \item We do not detect dark matter in the two Virgo UCDs with
        published dynamical masses and dynamical mass-to-light ratios,
        which happen to be the most massive of the Virgo UCDs
        measured, despite previous work indicating that the Virgo UCDs
        had higher mass-to-light ratios than the Fornax UCDs.
    \end{enumerate}

  \item The UCDs did not conform to a metallicity-luminosity relation,
    but did mostly lie above the metallicity-luminosity trend for
    early-type galaxies, which would be consistent with formation from
    present-day tidal stripping of dE,Ns.  Intermediate objects on
    either side of the data indicate that neither hypothesis can be
    ruled out in this way.

  \item The current data provide evidence for both formation
    hypotheses, and are insufficient to rule out either one.

\end{itemize}


\section*{Acknowledgments}
Based on observations obtained at the Gemini Observatory, which is
operated by the Association of Universities for Research in Astronomy,
Inc., under a cooperative agreement with the NSF on behalf of the
Gemini partnership: the National Science Foundation (United States),
the Science and Technology Facilities Council (United Kingdom), the
National Research Council (Canada), CONICYT (Chile), the Australian
Research Council (Australia), Ministario da Ciancia e Tecnologia
(Brazil) and Ministerio de Ciencia, Tecnologia e Innovacion Productiva
(Argentina). 

IC acknowledges the support from the grant of the president of the Russian Federation MD-3288.2012.2.

We thank Arna Karick for providing CCD photometry of the Fornax
Cluster objects. 

\bibliography{references}
\bibliographystyle{mn2e}

\bsp

\FloatBarrier

\begin{appendix} 
\section{Figures from the spectral fitting process} 
\label{ap-fig}
\FloatBarrier

\begin{figure}
\centering
\advance\leftskip-0.5cm
\epsfig{file=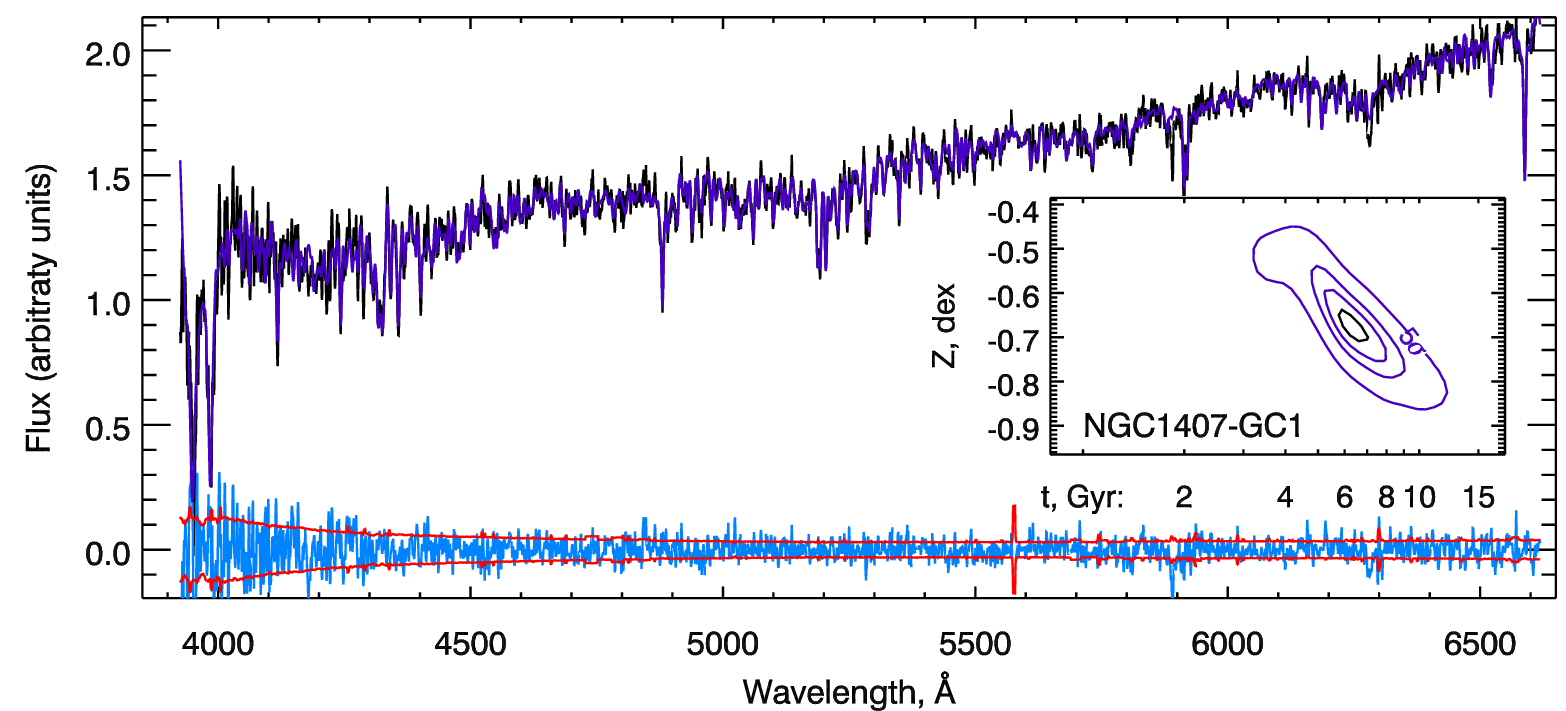,width=1.8\linewidth,angle=270,bb=
  64 360 677 643,clip=true} 
\label{fig-NGC1407GC1}
\caption{ {\sc nbursts} spectra, their best-fitting {\sc pegase.hr}
  templates \citep{LeBorgne2004}, fitting residuals and confidence
  levels of the age and metallicity determinations (inner panel) for
  the object NGC1407GC1. 
}
\end{figure}

\begin{figure*}
\noindent\makebox[\linewidth]{
\centering
\epsfig{file=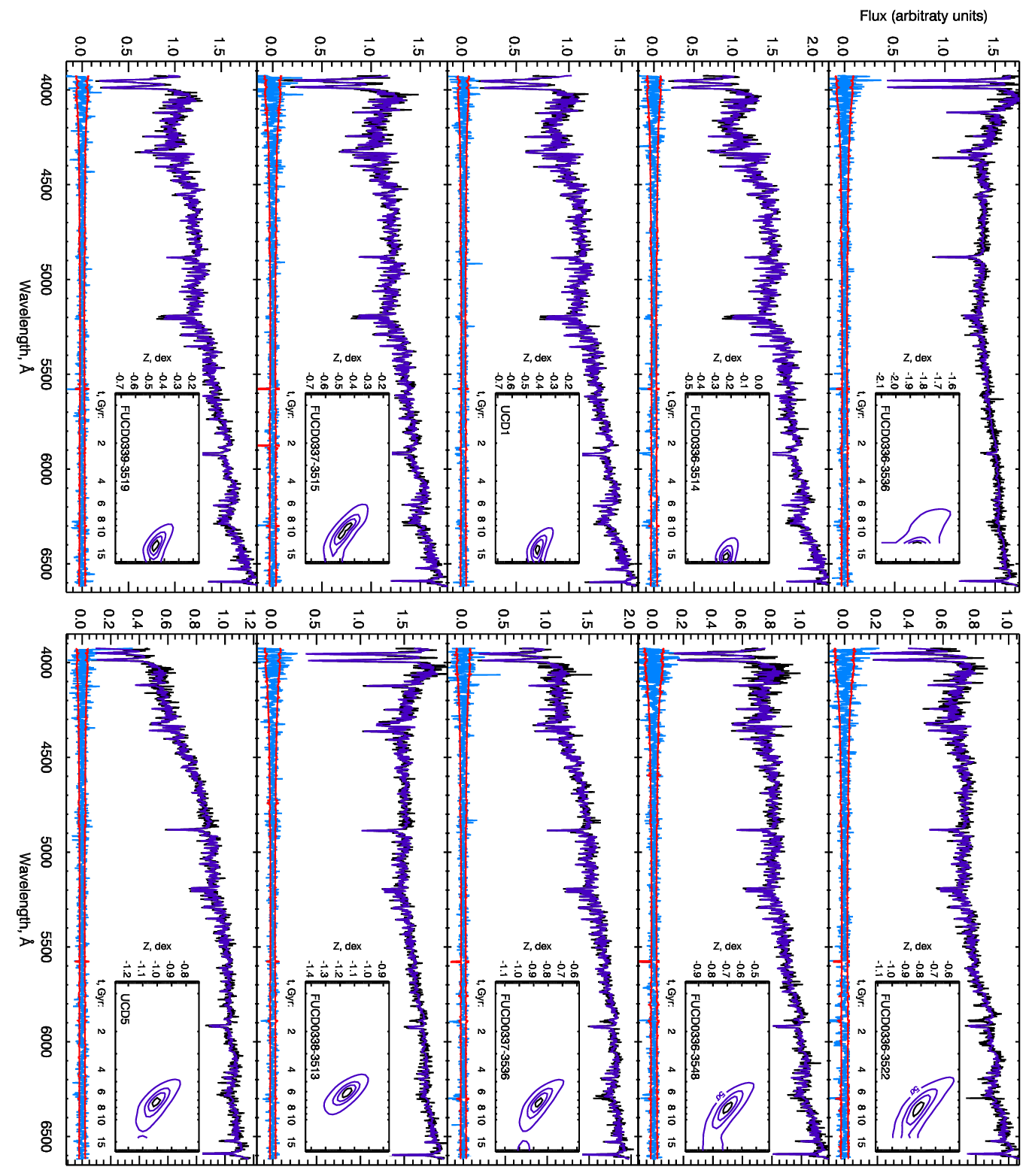,width=1.15\linewidth,angle=0}
}
\label{fig-FUCDs}
\caption{  {\sc nbursts} spectra, their best-fitting {\sc pegase.hr}
  templates \citep{LeBorgne2004}, fitting residuals and confidence
  levels of the age and metallicity determinations (inner panels)
  for the Fornax UCDs. 
}
\end{figure*}

\begin{figure*}
\noindent\makebox[\textwidth]{
\centering
\epsfig{file=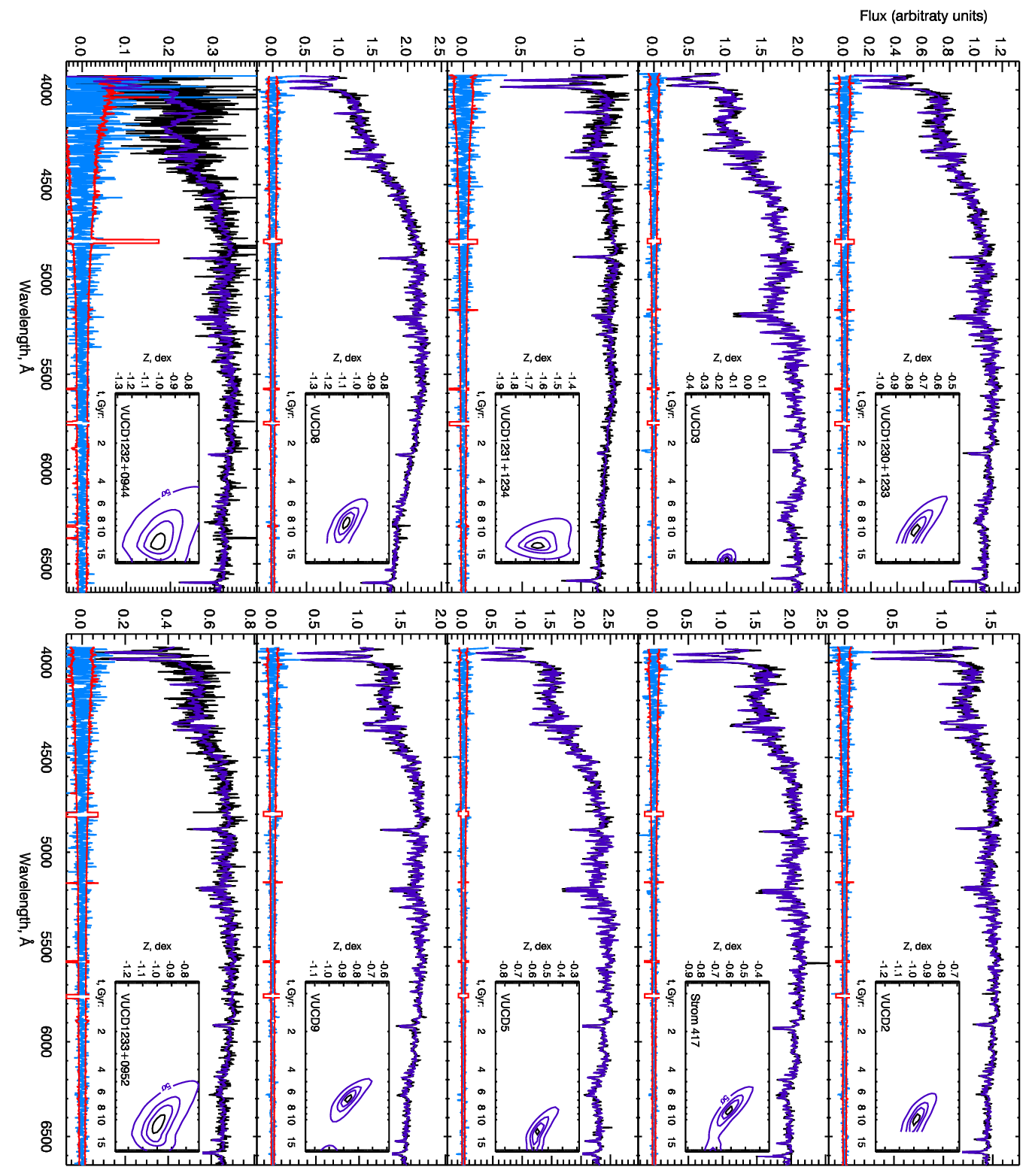,width=1.15\linewidth,angle=0}
}
\label{fig-VUCDs}
\caption{  {\sc nbursts} spectra, their best-fitting {\sc pegase.hr}
  templates \citep{LeBorgne2004}, fitting residuals and confidence
  levels of the age and metallicity determinations (inner panels) for
  the Virgo UCDs. 
}
\end{figure*}

\end{appendix}

\label{lastpage}

\end{document}